\documentclass[10pt,twocolumn,a4paper,showpacs,superscriptaddress]{revtex4}
\usepackage{epsfig,bm,dcolumn}
\usepackage{graphicx}

\begin{document}

\title{Evidence for the tensor meson exchange in the kaon photoproduction }

\author{Byung Geel Yu}%

\email{bgyu@kau.ac.kr}%
\affiliation{Research Institute of Basic Science, Korea Aerospace
University, Koyang, 412-791, Korea}

\author{Tae Keun Choi}%

\email{tkchoi@yonsei.ac.kr}%
\affiliation{Department of Physics, Yonsei University, Wonju,
220-710, Korea}

\author{W. Kim}%

\email{wooyoung@knu.ac.kr}%
\affiliation{Department of Physics, Kyungpook National University,
Daegu, 702-701, Korea}

\date{\today}

\begin{abstract}
The contribution of the tensor meson $K_2^*(1430)$ exchange in the
process $\gamma p\to K^+\Lambda(\Sigma^0)$ is investigated within
the Regge framework. Inclusion of the $K_2^*$ exchange in the
$K(494)+K^*(892)$ exchanges with the coupling constants chosen
from the SU(3) symmetry leads to a better description of the
production mechanism without referring to any fitting procedure.
This shows the significance of the role of the tensor meson
exchange to have the Regge theory basically free of parameters
with the SU(3) symmetry a good approximation for the meson-baryon
couplings.
\end{abstract}


\maketitle

        \section{Introduction}


It is believed that the Regge pole model could be an effective
theory for high-energy hadron reactions induced by electromagnetic
and mesonic probes \cite{word,storrow}.
The Regge models formulated in the $s$-channel helicity amplitude
(SCHA) \cite{walk} are favorable to the analysis of
photoproduction of pseudoscalar meson since they share essentially
the same production amplitude with that of the effective
Lagrangian approach except for the simple {\it reggeization} of
the $t$-channel meson poles \cite{levy,guid}. It is, therefore,
advantageous to work with the Regge poles in the SCHA in that one
exploits the effective Lagrangians to estimate the coupling
constants of the exchanged meson from the decay width or from the
symmetry consideration.
The application of these models to physical processes is, however,
limited by the large ambiguity in the coupling of meson trajectory
due to the fitting of the experimental data with few meson
exchanges. Within the framework of the $K+K^*$ Regge poles for
kaon photoproduction, to be specific, the coupling constants of
the $K^*$ to baryons were given too large as compared to those
either from the SU(3) symmetry prediction \cite{work} or from
other independent process such as the Nijmegen soft core potential
for the NN interaction \cite{nsc}. This large discrepancy, as
shown in Table \ref{cc1} below, demonstrates that the $K+K^*$
exchanges in current models are not enough to describe the process
up to $-t\approx 2$ GeV$^2$.

In this work we study the processes $\gamma p\to K^+\Lambda$ and
$\gamma p\to K^+\Sigma^0$ at forward angles within the Regge
framework and discuss the possibility of the model prediction
without fit parameters for the meson-baryon couplings. From our
previous analysis of the pion photoproduction \cite{bgyu}, we
recall, the inclusion of the tensor meson $a_2(1320)$ exchange in
the $\pi(140)+\rho(770)$ Regge poles led us to choose a rather
moderate value for the $\rho$-meson coupling constants for the
better description of the experimental data. (See the values
compared in Table \ref{cc1} below.) It is, then, natural to extend
the model of $K+K^*$ exchanges to obtain the parameter-free
prediction for the production mechanism by introducing the tensor
meson $K_2^*$.

      \section{tensor meson exchange at forward angles}

In the photoproduction amplitude for $\gamma(k)+ p(p)\to
K^+(q)+\Lambda(p')$,
\begin{eqnarray}\label{amp}
{\cal M}={\cal M}_K+{\cal M}_{K^*}+{\cal M}_{K_2^*}\,,
\end{eqnarray}
where the amplitudes relevant to the $K$ and $K^*$ Regge-pole
exchanges are given in Refs. \cite{guid,bgyu}, the exchange of the
$K^*_2$ Regge pole in the $t$-channel is written as
\cite{bgyu,giac,ysoh}
\begin{eqnarray}\label{tensor}
&&{\cal M}_{K_2^*}
=\bar{u}'(p')\varepsilon_{\alpha\beta\mu\nu}\,\epsilon^\mu k^\nu
q^\alpha q_\rho\Pi^{\beta\rho;\lambda\sigma}(q-k)\nonumber\\
&&\times\left[G^{(1)}_{K_2^*}(\gamma_\lambda
P_\sigma+\gamma_\sigma P_\lambda)+G^{(2)}_{K_2^*} P_\lambda
P_\sigma\right] {\cal
P}^{K_2^*}(s,t) u(p)
\end{eqnarray}
with the coupling constants $G^{(1)}_{K_2^*}$=$\frac{2g_{\gamma
KK^*_2}}{m^2_0}\frac{2g^{(1)}_{K^*_2NY}}{M}$,
$G^{(2)}_{K_2^*}$=$-\frac{2g_{\gamma
KK^*_2}}{m^2_0}\frac{4g^{(2)}_{{K^*_2}NY}}{M^2}$, and the momentum
$P$=$\frac{1}{2}(p+p')$. The mass parameter $m_0=1$ GeV is taken
for the dimensionless decay constant and $M$ is the nucleon mass.
The quantity
$\Pi_{\mu\nu;\rho\sigma}(q-k)=\frac{1}{2}(\eta_{\mu\rho}\eta_{\nu\sigma}
+\eta_{\mu\sigma}\eta_{\nu\rho})-\frac{1}{3}\eta_{\mu\nu}\eta_{\rho\sigma}
$ with $\eta_{\mu\nu}=-g^{\mu\nu} +
(q-k)^{\mu}(q-k)^{\nu}/m_{K^*_2}^2$ is the polarization tensor of
the $K^*_2$ meson.
\\

According to the duality expressed as the finite energy sum rule
between the $s$-channel resonances and the $t$-channel Regge poles
\cite{fesr},
\begin{eqnarray}\label{fesr}
\int_{s_0}^{\bar{s}}ds\,s^n\, {\rm Im\,}{\cal
A}_{res}(s,t)=\sum_{j=K^*,K_2^*,
\cdots}\gamma_j(t)\frac{\bar{s}^{\,\alpha_j+n+1}}{{\alpha_j}+n+1}\
,\label{fesr}
\end{eqnarray}
that the imaginary part of the resonance amplitude
does not vanish by the optical theorem in the left hand side of
Eq.(\ref{fesr}) is in effect equivalent to imply $\gamma_{K^*}\neq
\gamma_{K_2^*}$, $\cdots$, in the right hand side, i.e., the
violation of the exchange degeneracy (EXD) by the different
residues between the $K^*$ and $K_2^*$ in the leading $K^*$
trajectory \cite{storrow,word,bgyu}. This proves that the weak EXD
of the pair $K^*$-$K_2^*$ is a good approximation, and hence, both
the two contribute independently with the different residues
(different coupling vertices in the present scheme), but share the
same phase of the signature factor
with each other. Thus, we use the $K_2^*$ Regge pole of the spin-2
\begin{eqnarray}\label{pi-regge}
&&{\cal
P}^{K_2^*}(s,t)=\frac{\pi\alpha'_{K_2^*}}{\Gamma(\alpha_{K_2^*}(t)-1)}
\frac{e^{-i\pi\alpha_{K_2^*}(t)}}{\sin\pi\alpha_{K_2^*}(t)}
\left(\frac{s}{s_0}\right)^{\alpha_{K_2^*}(t)-2}
\end{eqnarray}
with the rotating phase for the nonzero imaginary part of the
amplitude. Here the trajectory
\begin{eqnarray}\label{tensor-traj}
\alpha_{K_2^*}(t)=0.83\,(t-m^2_{K^*_2})+2
\end{eqnarray}
is taken for the $K_2^*$ with the slope the same as that of the
$K^*$  \cite{guid} and the scale factor $s_0$ is chosen as $1$
GeV$^2$.

Avoiding fit parameters for the coupling constants of all
exchanged mesons considered here we use the SU(3) relations to
determine their values. We begin with the estimate of the $K^*NY$
coupling, while the relatively well-established coupling constant
$g_{KNY}$ and radiative decay constant $g_{\gamma K^\pm
K^*}=0.254$ are taken the same as those in Ref. \cite{guid} for
comparison.
We estimate the coupling constants of the vector meson
$g^{v(t)}_{K^*NY}$ by using the SU(3) relations in which case
$g^v_{\rho NN}=2.6$ is taken from the universality of $\rho$ meson
coupling with the ratio $\alpha^v=1$. For the tensor coupling of
the $\rho$ meson, $g^t_{\rho NN}$, we use $\kappa_\rho=6.2$ with
the ratio $\alpha^t=0.4$ from the SU(6) quark model prediction
\cite{work}.

The radiative decay, $K^*_2\to \gamma K$, is empirically known and
the width reported in the Particle Data Group is,
$\Gamma_{K^*_2\to K\gamma}=0.24\pm 0.05$ MeV. The decay width
corresponding to the $K^*_2 K\gamma$ vertex in Eq.(\ref{tensor})
is given by \cite{giac}
\begin{eqnarray}\label{ten-decay}
\Gamma_{K^*_2\to K\gamma}=\frac{1}{10\pi}\, \left(\frac{g_{\gamma
KK^*_2}}{m^2_0}\right)^2
\left(\frac{m_{K^*_2}^2-m^2_K}{2m_{K^*_2}}\right)^5\,,
\end{eqnarray}
from which $g_{\gamma K K^*_2}=0.276$ is obtained.
Since there are no informations currently available for the
$K_2^*NY$ couplings except for those $a_2NN$ and $f_2(1270)NN$, we
resume the SU(3) symmetry for the tensor meson nonet coupling to
baryons where the $K^*_2NY$ coupling constants are given by
\begin{eqnarray}\label{tensor-k}
&&g^{(1,2)}_{K^*_2N\Lambda}=-\frac{1}{\sqrt{3}}(1+2\alpha_{(1,2)})
g^{(1,2)}_{a_2
NN}\,,\nonumber\\
&&g^{(1,2)}_{K^*_2N\Sigma}=(1-2\alpha_{(1,2)})g^{(1,2)}_{a_2
NN}\,,
\end{eqnarray}
and estimate the $K_2^*NY$ coupling constants from the knowledge
of the $a_2NN$ couplings in existing estimates.  In order for the
above SU(3) predictions to be reliable, it is, therefore, of
importance to choose the $a_2NN$ coupling constants on the firm
ground as well as the ratio $F/D$.
%
\begin{table}{}
\caption{SU(3) predictions for the coupling constants of the
$a_2NN$, and $K_2^*NY$ from the given $f_2NN$ coupling constant. }
\label{tb5}
\begin{tabular}{ccccccl}
                              &  A      &     B        &  C &$(\frac{F}{D})_{\rm exp}=-1.8\pm 0.2$    &\\
\hline\hline
$g_{f_2NN}^{(1)}$             & 3.38$^a$&5.26$^b$      &6.45$^c$    &       &\\%
\hline
$g_{a_2NN}^{(1)}$              &  0.6   & 0.94  &1.15   & $\alpha_{(1)}=2.67$, $\frac{F}{D}=-1.6$& \\%
$g^{(1)}_{K_2^{*}p\Lambda}$    & -2.20  &-3.44  & -4.21 &                 &\\%
$g^{(1)}_{K_2^{*}p\Sigma^0}$   & -2.60  &-4.08  & -4.99 &                 &\\%
\hline
$g_{a_2NN}^{(1)}$              &  0.73  & 1.14  & 1.4   & $\alpha_{(1)}=2.25$, $\frac{F}{D}=-1.8$&      \\%
$g^{(1)}_{K_2^{*}p\Lambda}$    & -2.32  & -3.62 & -4.45 &                 &\\%
$g^{(1)}_{K_2^{*}p\Sigma^0}$   & -2.56  & -3.99 & -4.9  &                 &\\%
\hline
$g_{a_2NN}^{(1)}$              &  0.84  & 1.3   & 1.6   & $\alpha_{(1)}=2.0$, $\frac{F}{D}=-2.0$&      \\%
$g^{(1)}_{K_2^{*}p\Lambda}$    & -2.42  & -3.75 & -4.62 &                &\\%
$g^{(1)}_{K_2^{*}p\Sigma^0}$   & -2.52  & -3.9  & -4.8  &                &\\%
\hline
\end{tabular}
\end{table}
For verification we will check the consistency of the chosen
$a_2NN$ coupling constants  by using the SU(3) relation
\begin{eqnarray}\label{su3}
g_{f_2NN}^{(1,2)}=\frac{1}{\sqrt{3}}(4\alpha_{(1,2)}-1)\,g_{a_2
NN}^{(1,2)}\,,
\end{eqnarray}
with the ratio and the $f_2NN$ coupling constants which were given
in more detail in the literature \cite{renn,gold,borie,engels}.

Based on the dispersion relation and on the tensor meson dominance
(TMD) \cite{renn} the $f_2NN$ coupling constants were investigated
in the analysis of the backward $\pi N$ scattering \cite{gold} and
the $\pi\pi\to N\bar{N}$ partial-wave amplitudes
\cite{borie,engels}.
In these analyses we first note that $g^{(2)}_{f_2NN}\approx0$ was
obtained in common and we adopt this in Eq. (\ref{su3}) together
with $g^{(2)}_{a_2NN}\approx 0$ in accordance with our previous
result \cite{bgyu}. Therefore, it is reasonable to assume that
$g^{(2)}_{K^*_2NY}$ is small enough to be neglected in Eq.
(\ref{tensor-k}). We now focus on the estimate of
$g^{(1)}_{f_2NN}$ coupling constants from these analyses to
summarize the results in the first raw of Table \ref{tb5}. (In the
convention of Refs. \cite{renn,gold,engels,klei,klei1},
$G^{(1,2)}_{f_2NN}=M\gamma^{(1,2)}_{f_2NN}=4g^{(1,2)}_{f_2NN}$ in
Eq. (\ref{tensor}) and
$G_{f_2\pi\pi}=2m_{f_2}\gamma_{f_2\pi\pi}=4g_{f_2\pi\pi}$ in Eq.
(\ref{tmd}) below.) The value with the superscript $a$ in the
column A is obtained from the quantity
$\gamma_{f_2\pi\pi}\gamma^{(1)}_{f_2NN}/4\pi=10.4$ GeV$^{-2}$
which was extracted from the $\pi N$ scattering \cite{gold}. In
the column B the value with the $b$ is from $g^{(1)\
2}_{f_2NN}/4\pi=2.2\pm0.9$ which was obtained in the dispersion
analysis of the $\pi\pi\to N\bar{N}$ \cite{borie}. The value with
the $c$ in the last column is from $G^{(1)\
2}_{f_2NN}/4\pi=53\pm10$ using the Regge model for the backward
$\pi N$ scattering \cite{engels}, which also agrees with that
obtained from other independent processes \cite{klei,klei1}. In
each column in Table \ref{tb5} we display the SU(3) predictions
from Eqs. (\ref{tensor-k}) and (\ref{su3}) for the
$g^{(1)}_{a_2NN}$ and $g^{(1)}_{K^*_2NY}$ with the ratio
$F/D=-1.8\pm 0.2$, which was determined to agree with the
Regge-pole fit to the high energy experiments based on the SU(3)
symmetry for the residues of the tensor meson nonet coupling to
baryons \cite{sarma,gross}. On the other hand, we find that, among
these values, the choices of $g_{f_2NN}^{(1)}=6.45$ and
$g_{a_2NN}^{(1)}=1.4$ or $1.6$ in the column C with
the ratio $\alpha_{(1)}=2.0$ or 2.25 are in fair agreement with
$G^{(1)\ 2}_{a_2NN}/4\pi\approx3$
and $|G^{(1)}_{a_2NN}|\approx 6$
obtained from the analyses of pion photoproduction \cite{klei} and
the Compton scattering \cite{klei1}, respectively.
Thus, we favor to choose the SU(3) value $g^{(1)}_{a_2NN}=1.4$ as
a median value together with $g^{(1)}_{f_2NN}=6.45$ for the
estimate of the $g^{(1)}_{K_2^*NY}$ in Eq.(\ref{tensor-k}).

\begin{table}[t]
\caption{\label{cc1} Meson-baryon coupling constants for the
exchanged mesons in the $\gamma p\to \pi^+n$ \cite{bgyu} and
$\gamma p\to K^+\Lambda(\Sigma^0)$ processes. The models LMR and
GLV refer to Refs. \cite{levy,guid}, respectively. An overall
factor $\lambda=2.18$ is taken for the absorption correction in
the LMR model.  }
\begin{tabular}{lcccccl}
                             &NSC97a  & LMR       &  GLV       &Present work  \\
\hline
$g_{\pi NN}/\sqrt{4\pi}$      &3.71        &3.82       &  3.81      & 3.81    \\%
$g^v_{\rho NN} $              &   2.97      &2.8        & 3.4       & 2.6      \\%
$g^t_{\rho NN}$               &   12.52      &40.88      & 20.74      &  16.12  \\%
$g^{(1)}_{a_2NN}(g^{(2)}_{a_2NN})$&      &-          & -          &1.4 (0)  \\%
\hline
$g_{Kp\Lambda}/\sqrt{4\pi}$    &  -3.82      &-3.87      &  -3.26     &-3.26    \\%
$g_{Kp\Sigma^0}/\sqrt{4\pi}$   &1.16         &0.76       &  1.26      & 1.26    \\%
%
$g^v_{K^*p\Lambda}$           &   -4.26      &-7.29$\lambda$ & -23    &  -4.5   \\%
$g^t_{K^*p\Lambda}$           &   -11.31     &-31.72$\lambda$& 57.5   &  -16.7  \\%
$g^v_{K^*p\Sigma^0}$          &   -2.46     &-7.02$\lambda$ & -25     &-2.6     \\%
$g^t_{K^*p\Sigma^0}$          &   1.15       &26.82$\lambda$& 25      &3.2     \\%
$g^{(1)}_{K_2^{*}p\Lambda}(g^{(2)}_{K_2^{*}p\Lambda})$&&  -           & -   &-4.45 (0)\\%
$g^{(1)}_{K_2^{*}p\Sigma^0}(g^{(2)}_{K_2^{*}p\Sigma^0})$&&  -         & -   &-4.9 (0) \\%
\hline
\end{tabular}
\end{table}
%


We present in Table \ref{cc1} the meson-baryon coupling constants
of the exchanged mesons in the Regge models for the pion and kaon
photoproduction. The corresponding values from Nijmegen soft core
potential (NSC97a) is listed for comparison \cite{nsc}. The
pseudoscalar meson coupling constants in the NSC97a are deduced by
using the proportional expressions of the given pseudovector ones
in Ref. \cite{esc04}. Note that the $K^*NY$ coupling constants
determined from the SU(3) relations in the present work are the
same order of the magnitude with those obtained from the NSC97a.
The tensor meson couplings are obtained from SU(3) relations with
$\alpha_{(1)}=2.25$.

Before closing this section let us comment on the TMD in relation
with the determination of the $f_2NN$ coupling constants
\cite{renn,bgyu}. The TMD with the $f_2$-pole dominance in the
$\pi N$ scattering process leads to the following identity,
\begin{eqnarray}\label{tmd}
\frac{2}{M}(g^{(1)}_{f_2NN}+g^{(2)}_{f_2NN})=\frac{g_{f_2\pi\pi
}}{m_{f_2}}
\end{eqnarray}
which estimates $g^{(1)}_{f_2NN}=2.13$ and $g^{(2)}_{f_2NN}=0$
with the known coupling constant $g_{f_2\pi\pi}=5.76$. The
coupling constant $g^{(1)}_{f_2NN}$ predicted by the TMD is small
and inconsistent with those discussed above. Since the validity of
the TMD in such a simple $f_2$-pole description is questionable
and needs further test \cite{raman,suzuki}, we disregard the TMD
prediction in this work, though a viable hypothesis analogous to
the VMD.

           \section{Results and discussion}

    \begin{figure}[t]
    \begin{center}
    \includegraphics[width=0.95\linewidth]{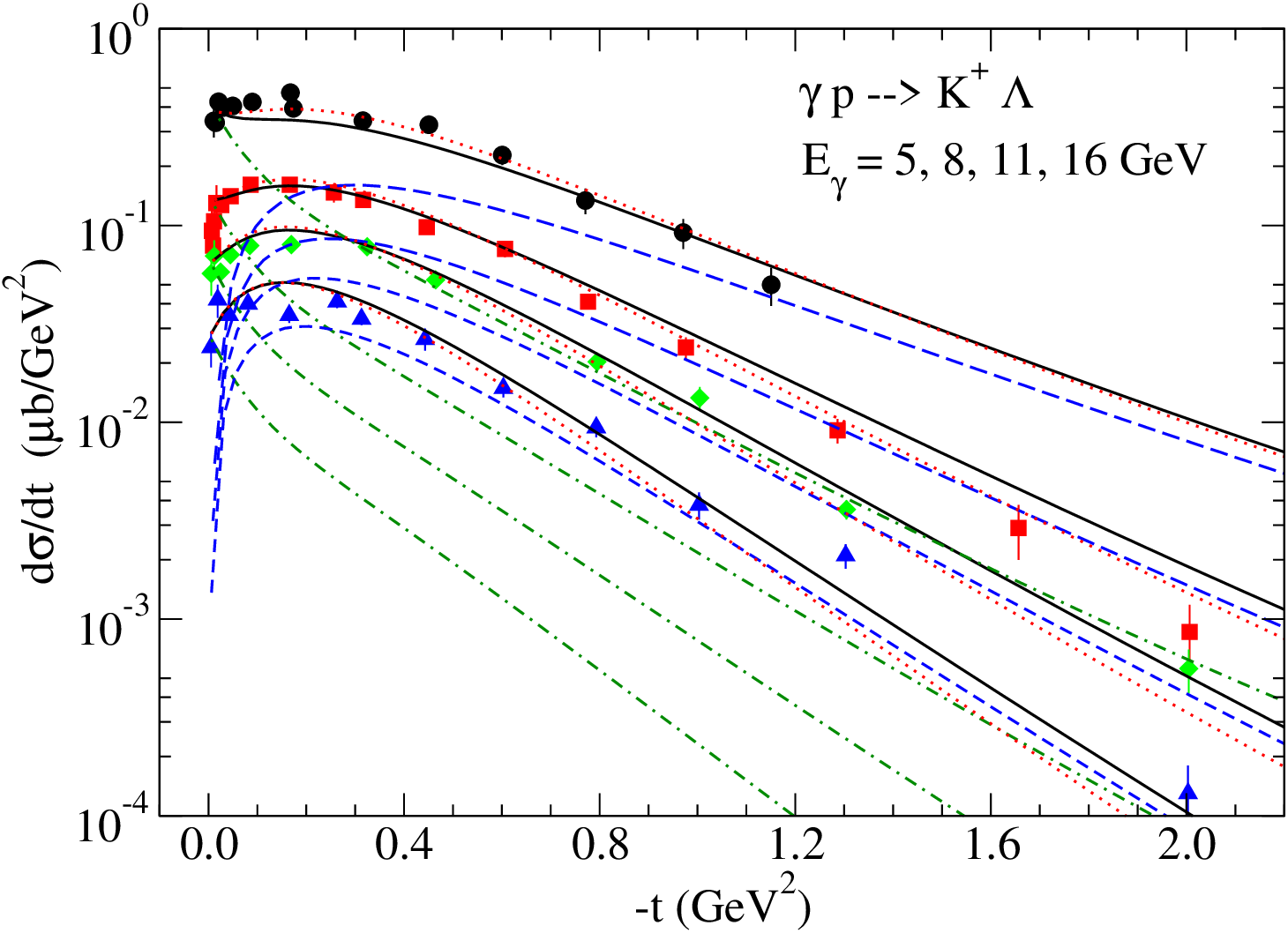}
    \end{center}
    \caption{(Color online) Differential cross sections
$\frac{d\sigma}{dt}$ for $\gamma p\to K^+\Lambda$ at photon
energies $E_\gamma=5,\,8,\,11,\,16$ GeV, respectively. Solid lines
(black) result from the gauge invariant $K+K^*+K^*_2$ exchanges in
the present model. Dash-dotted lines (green) represent the $K+K^*$
exchanges in the present model. Dashed ones (blue) denote the
$K^*_2$ contributions.  Dotted lines (red) are from the GLV model.
Data are taken from Ref.\cite{boyar}.}\label{fig:fig1}
    \end{figure}

\begin{figure}[]
\begin{center}
\includegraphics[width=0.95\linewidth]{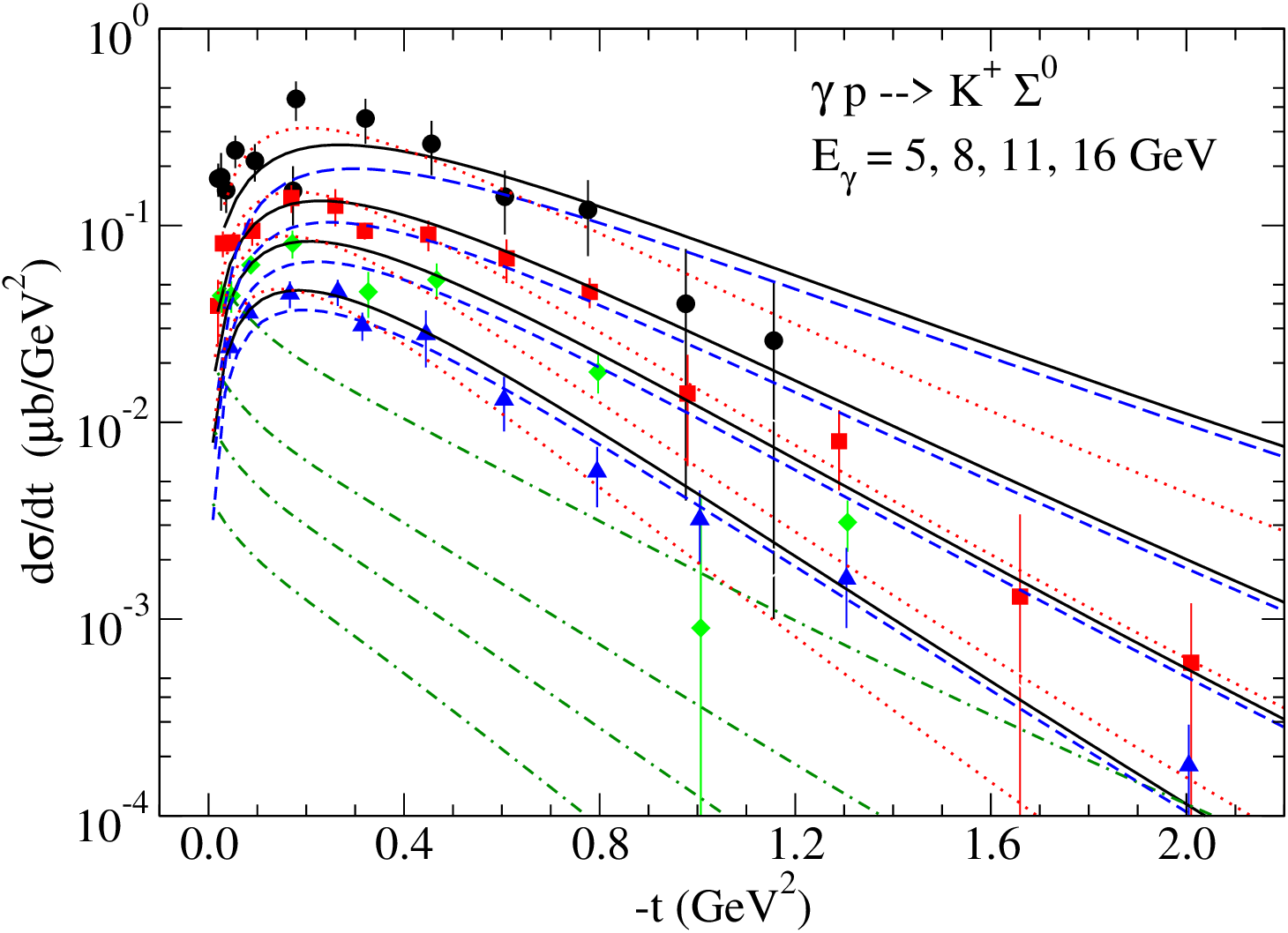}
\end{center}
\caption[]{(Color online) Differential cross sections
$\frac{d\sigma}{dt}$ for $\gamma p\to K^+\Sigma^0$ at photon
energies $E_\gamma=5,\,8,\,11,\,16$ GeV, respectively. Notations
are the same with Fig.\ref{fig:fig1}. Data are taken from Ref.
\cite{quinn}.} \label{fig:fig2}
\end{figure}

Figures \ref{fig:fig1} and \ref{fig:fig2} show the differential
cross sections for $\gamma p\to K^+\Lambda$ and $\gamma p\to
K^+\Sigma^0$ at photon energies $E_\gamma=5,\,8,\,11$, and $16$
GeV, respectively. It is clear that the $K+K^*$ exchanges with the
SU(3) coupling constants (the green dash-dotted line) can hardly
reproduce the cross section at any photon energy but the $K_2^*$
exchange replaces the role that has been attributed to the $K^*$
in Refs. \cite{levy,guid}, instead. This feature of the production
mechanism should be different from that of the $K+K^*$ exchanges
(the red dotted lines) in the GLV model, even if it yields the
cross sections comparable to the solid ones with very large $K^*$
coupling constants as shown in Table \ref{cc1}. This tendency
continues to the $\gamma p\to K^+\Sigma^0$ case, though the cross
section in Fig. \ref{fig:fig2} is in less agreement with data at
the photon energy $E_\gamma=5$ GeV due to the small couplings of
$KN\Sigma$ and $K^*N\Sigma$. In conclusion, the features of the
production mechanism in the present work result from the $K+K_2^*$
exchanges, but not from those of the $K+K^*$ as described in
previous studies. In both processes the $K^*_2$ interferes
constructively with the sum total of $K+K^*$ to reproduce the
solid line. To a change of the $K^*_2$ coupling constant within
the uncertainty of the $F/D$ ratio, the cross section shows
sensitivity to some degree. But in any cases we find that the
$K^*_2$ plays the key role to reproduce the whole structure of the
cross section.

\begin{figure}[]
\begin{center}
\includegraphics[width=0.95\linewidth]{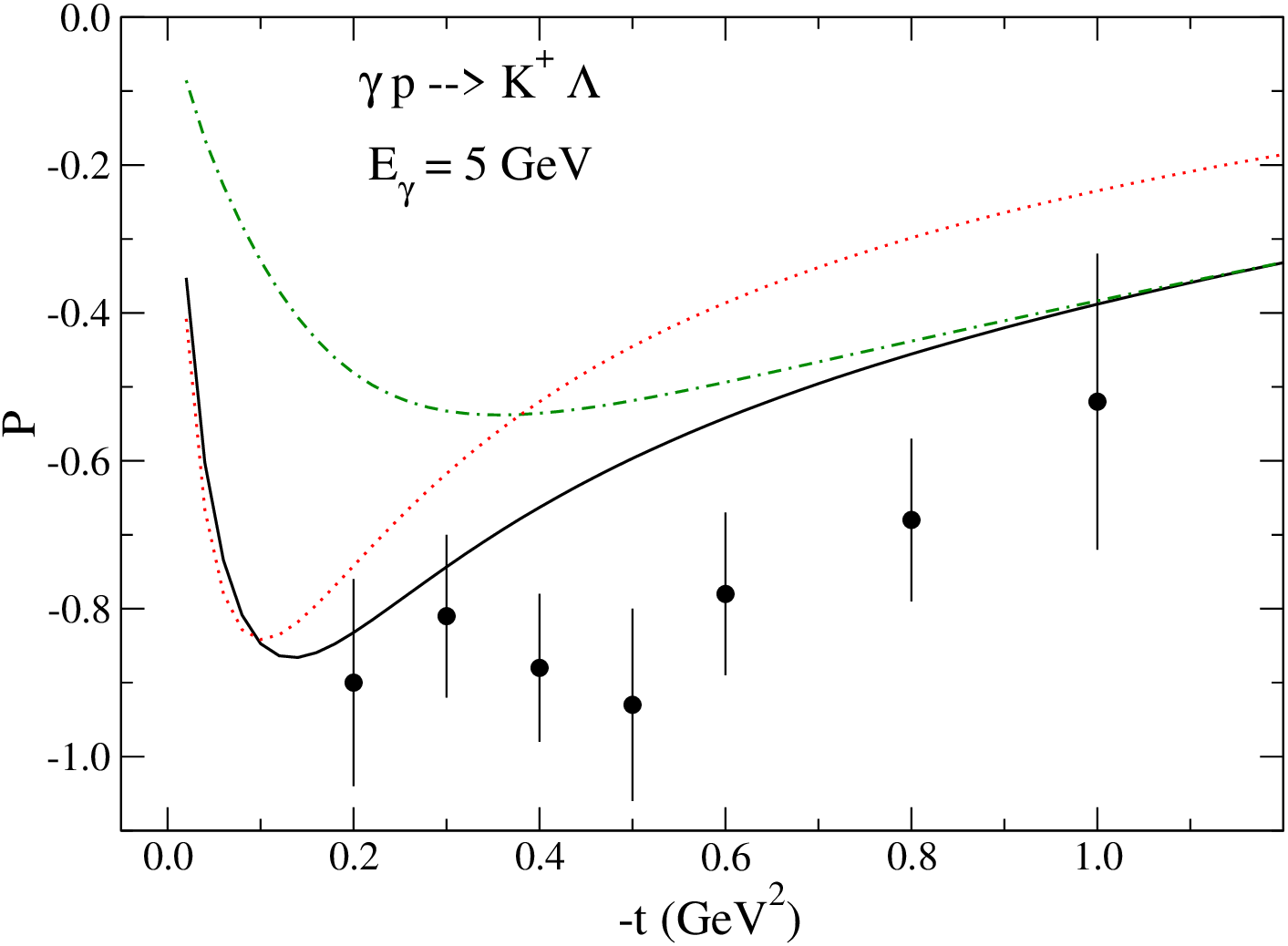}
\end{center}
\caption[]{(Color online) Recoil polarization asymmetry for
$\gamma p\to K^+\Lambda$ at $E_\gamma=5$ GeV. Notations are the
same with Fig.\ref{fig:fig1}. Data are taken from
Ref.\cite{vogel}.} \label{fig:fig3}
\end{figure}

The recoil polarization $P$ is analyzed in Fig. \ref{fig:fig3}.
The negative value of the $P$ observed in the experiment indicates
a spin-down of the recoiled $\Lambda$, supporting our SU(3)
predictions for the negative signs of the $K^*_2NY$ and $K^*NY$
couplings as well. Note that the inclusion of the $K_2^*$ makes
improved the model prediction from that of $K+K^*$ to the
experimental data closely.
For the photon polarization in the $\gamma p\to K^+\Lambda$, we
obtain exactly the same result at $E_\gamma=16$ GeV as presented
in Ref. \cite{guid} which shows the rapid approach to unity by the
dominance of the natural parity exchanges, $K^*+K^*_2$ over the
unnatural parity $K$.

\begin{figure}[b]
\begin{center}
\includegraphics[width=0.95\linewidth]{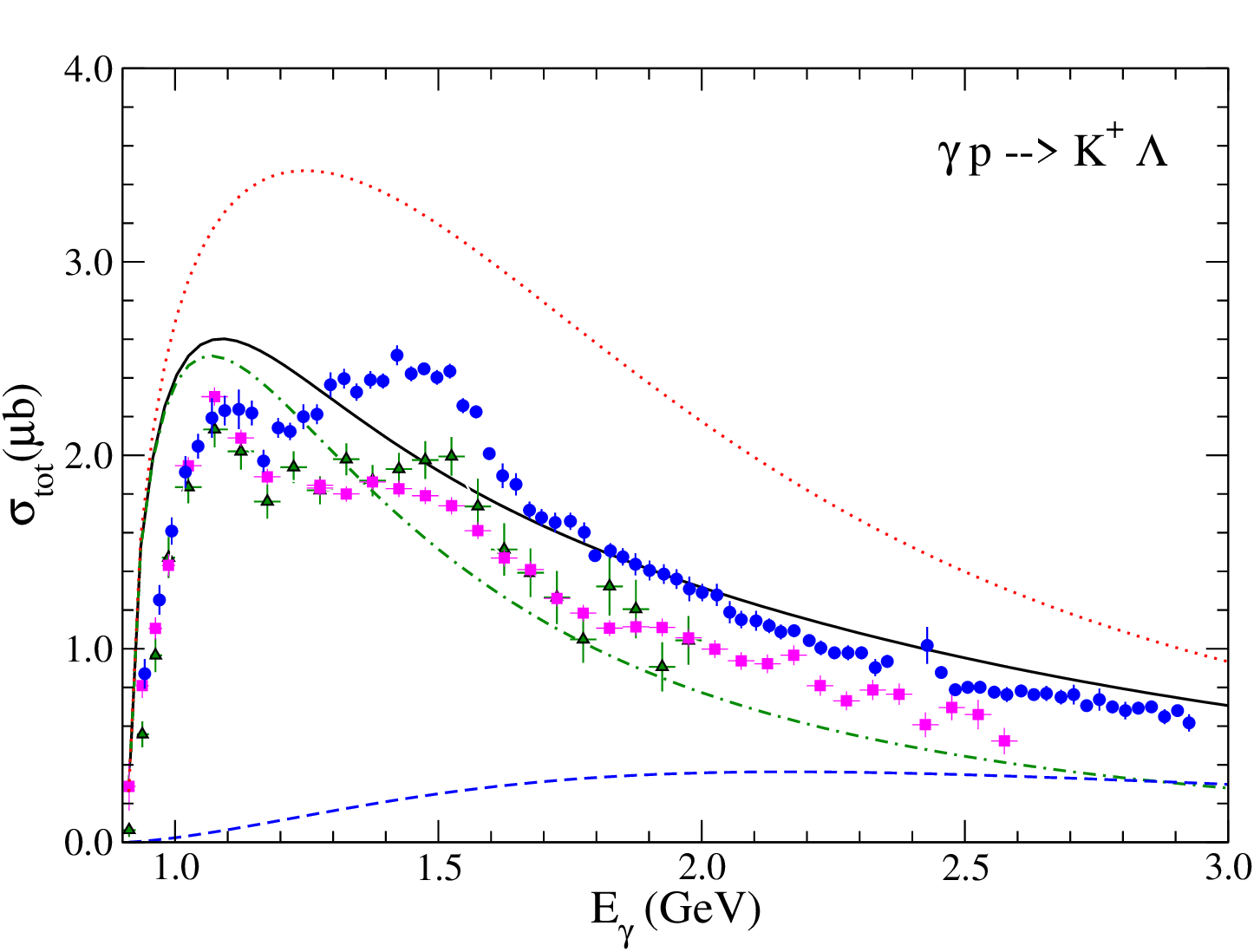}
\end{center}
\caption{(Color online) Total cross section for $\gamma p\to
K^+\Lambda$ up to $E_\gamma=3$ GeV. Notations are the same with
Fig.\ref{fig:fig1}. Data are taken from
\cite{saph98,saph03,clas05}.} \label{fig:fig4}
\end{figure}

Finally, we should remark upon the effect of the $K_2^*$ exchange
on the lower energy region. Figure \ref{fig:fig4} shows the total
cross section measured at the SAPHIR/ELSA \cite{saph98,saph03} and
the CLAS/JLab experiments in the resonance region \cite{clas05}.
The size of the cross section largely depends on the magnitude of
the leading coupling constant $g_{KN\Lambda}$, as can be expected
from the significance of the nucleon Born term in this region. The
destructive interference between the $K$ and $K^*$ exchange leads
to a sizable reduction of the total cross section, while the
$K_2^*$ gives the additive contribution to the $K+K^*$, and we
obtain a good agreement with the experimental data by using the
same $g_{KN\Lambda}$ as that of the GLV model.
It is understood that the overestimation of the cross section (the
red dotted line) by the latter model is, therefore, another
evidence for the inadequacy of such a large $K^*$ coupling
constants as fitted to the high-energy data.

In this letter, with such compelling evidences as shown, we have
clarified two points that have been obscure as concerns the Regge
approach to kaon photoproduction based on the $s$-channel helicity
amplitude \cite{levy,guid,cort}; one is our current
misunderstanding of the large $K^*$ contribution due to the
fitting procedure without the $K^*_2$. The other is the
possibility of the Regge theory to be basically free of parameters
with the SU(3) symmetry quite a good approximation for the
meson-baryon couplings by considering the tensor meson $K^*_2$.


        \section*{Acknowledgments}
This work was supported by Basic Science Research Program through
the National Research Foundation of Korea(NRF) funded by the
Ministry of Education, Science and Technology(2010-0013279).

\end{document}